\documentclass[12pt,a4paper]{article}
\usepackage[utf8]{inputenc}
\usepackage[T1]{fontenc}
\usepackage{amsmath}
\usepackage{amsfonts}
\usepackage{amssymb}
\usepackage{graphicx}
\usepackage{authblk}
\usepackage{mathrsfs}
\newcommand{\be}{\begin{equation}}
\newcommand{\ee}{\end{equation}}
\newcommand{\bea}{\begin{eqnarray}}
\newcommand{\eea}{\end{eqnarray}}

\renewcommand{\baselinestretch}{1.5}
\begin{document}

\title{\vspace{-1.8in} Effective field theory for closed strings \\ near the Hagedorn temperature }
\author{Ram Brustein, Yoav Zigdon}
\affil{{\normalsize Department of Physics, Ben-Gurion University, Beer-Sheva 84105, Israel} \\
{\small\tt ramyb@bgu.ac.il\ \ \ \ yoavzig@post.bgu.ac.il}}
\date{}
\maketitle
\begin{abstract}
We discuss interacting, closed, bosonic and superstrings in thermal equilibrium at temperatures close to the Hagedorn temperature in flat space. We calculate S-matrix elements of the strings at the Hagedorn temperature and use them to construct a low-energy effective action for interacting strings near the Hagedorn temperature. We show, in particular, that the four-point amplitude of massless winding modes leads to a positive quartic interaction. Furthermore, the effective field theory has a generalized conformal structure, namely, it is conformally invariant when the temperature is assigned an appropriate scaling dimension. Then, we show that the equations of motion resulting from the effective action possess a winding-mode-condensate background solution above the Hagedorn temperature and present a worldsheet conformal field theory, similar to a Sine-Gordon theory, that corresponds to this solution. We find that the Hagedorn phase transition in our setup is  second order, in contrast to a first-order transition that was found previously in different setups.
\end{abstract}
\newpage
\renewcommand{\baselinestretch}{1.5}\normalsize
	\tableofcontents

	\section{Introduction}

String theory can be studied at finite temperatures by considering a target space with $d$ flat non-compact dimensions and one dimension compactified on a circle with periodic boundary conditions for  bosons and antiperiodic boundary conditions for fermions \cite{Polchinski1},\cite{Rohm},\cite{McGuigan}. When the temperature approaches the ``Hagedorn temperature'' \cite{Hagedorn}, which is inversely proportional to the string length, the one-loop free energy of free strings naively diverges \cite{fubinivenezia,Weinberg}.  Initially, it was thought that the Hagedorn temperature is a maximal, or a limiting temperature \cite{Sundborg1984,Tye1985}. However, the fact that at this temperature certain winding modes become massless, motivated the point of view that a phase transition takes place \cite{Sathiapalan1986,OBrien1987,AW}.  We refer to this winding mode as the ``thermal scalar.''

Atick and Witten \cite{AW} argued that the tree-level coupling between the dilaton and two thermal scalars leads to a first-order transition, at a critical temperature below the Hagedorn temperature. However, Horowitz and Polchinski (HP) \cite{HP} pointed out that at temperatures close to the Hagedorn temperature, the coupling between the dilaton and two thermal scalars is suppressed by either derivatives or by a factor of the small deviation from the Hagedorn temperature. They therefore argued that the dilaton coupling is not the dominant tree-level coupling and consequently does not determine the order of the phase transition. The case that HP considered, aiming to relate this setup to a black hole \cite{HP1,Damour:1999aw}, is that of a single isolated string with fixed energy.  They argued that, in this case the dominant interaction is between two thermal scalars and the graviton associated with the compact dimension (the ``radion'') and concluded that the transition is first order. This conclusion was accepted in subsequent discussions in the literature (see, for example, \cite{Gubser},\cite{Barbon1},\cite{Mertens:2014dia}).

We wish to describe the physics of closed strings propagating in flat space in the Hagedorn phase, when the temperature is close to, either below or above the Hagedorn temperature, taking into account string interactions. While HP considered a single isolated string, we wish to consider a different situation in which the string is coupled to a translation-invariant thermal bath of strings at a temperature close to the Hagedorn temperature. The bath fixes the temperature of the string, or the size of the thermal circle, forcing the radion to vanish. If, in addition, the string coupling is finite and small, then the tree-level thermal scalar quartic self-interaction is the dominant interaction.

Our description of the Hagedorn phase relies on a Euclidean target space low-energy effective field theory (EFT) for the thermal scalar, graviton and dilaton.  We determine the action of the EFT by matching the S-matrix elements of string theory and the EFT, a method used frequently in string theory literature \cite{Effective0,Effective1,Effective2}.   In this paper, we calculate the coefficient of the thermal scalar quartic self-interaction term and, in particular, show that it is positive.  We then write a tree-level EFT action for the light modes in the vicinity of the Hagedorn temperature. We find a simple background solution for which the metric is flat and the rest of the light fields are set to constants. In particular, the thermal scalar condenses above the Hagedorn temperature.  A world-sheet (WS) conformal field theory (CFT) similar to Sine-Gordon is found and used to calculate correlation functions in conformal perturbation theory which correspond to the S-matrix elements of the EFT. The CFT analysis verifies the validity of the target space EFT for our setup.

Furthermore, it is demonstrated that the target space  EFT possesses a generalized conformal structure \cite{Jevicki1,Jevicki2,Skenderis}, that is, it is  conformally invariant when the temperature varies under the conformal transformation. Typically, field theories at finite temperatures are not conformal, because the temperature is treated as a scale that does not transform under Weyl transformations. However, by choosing to transform the temperature that depends on the metric compact-compact component, we show how conformal symmetry comes about. The generalized conformal structure is sufficient to ensure that the correlation functions obey the same properties as a truly conformally invariant theory. Our results confirm and extend the qualitative analysis in \cite{dSHigh,CFT4dS} and fix the string-theory-dependent coefficients of the EFT.

When the strings are coupled to a translation-invariant thermal bath, the thermal scalar condensate solution is a thermal state that is known to be one  possessing maximally allowed  pressure, namely, a pressure equal to the energy density  \cite {AW}. This follows from the free energy scaling as the square of the temperature in any dimension. Furthermore, this state is a state of maximal entropy \cite{saskatoon,dSHigh,CFT4dS}. A remarkable fact is that the symmetries and causal structure  of the thermal scalar condensate solution and of dS space are identical. The strings are described by a conformally invariant effective theory, while for dS space, the isometries are those of  the conformal group. For further discussion of the implications, see \cite{CFT4dS}.

We then discuss the nature of the Hagedorn phase transition for type II superstring theory and find that in this case the transition is second order.
It is amusing to note that  for the bosonic closed string, there are two additional momentum modes that become massless at the Hagedorn temperature. The quartic coupling of these modes is identical to the quartic coupling of the thermal scalar. These modes condense below the Hagedorn temperature. What is special about the momentum modes is that, if one ignores the problems with the closed string tachyon, then the nature of the Hagedorn transition is changed in comparison to type II strings, becoming a smooth crossover as we will show.

Previous discussions of  winding mode condensates and WS CFTs that describe them, include, for example, Maggiore \cite{Maggiore}, who introduced a WS CFT and argued that above the Hagedorn temperature, winding modes condense, giving rise to discrete values for temperature and non-commutative spatial geometry described by a quantum algebra. The authors of \cite{Sunny1},\cite{Sunny2} considered a similar Sine-Liouville theory,  motivated, at least in part, by its correspondence to black holes. While we consider a fixed circumference  thermal circle and a constant dilaton, in \cite{Sunny2} a background with linear dilaton and a cigar shaped space is considered.  For additional discussions of winding mode condensates, see \cite{Silverstein1,Silverstein2,Silverstein3} and \cite{Headrick} for a review and additional references.

The paper is organized as follows. In Section 2, winding modes are reviewed and their string S-matrix elements are calculated. The effective action that reproduces these S-matrix elements is discussed in Section 3. In Section 4  we describe the  background solution and the corresponding WS CFT and in Section 5 we discuss the generalized conformal structure of the EFT.  The nature of the Hagedorn transition in is discussed in Section 6. We present our conclusion and discuss our interpretation of the results in Section 7. \newpage

\section{Thermal scalar vertices}

In this section we calculate the interaction terms in the EFT of the thermal scalar. In particular, we perform an explicit calculation of the four-point thermal scalar $(\phi \phi^*)^2$ interaction.  As far as we know, this specific calculation has not appeared in the literature. Related calculations can be found in \cite{Suyama} and \cite{DineKlebanov}. To calculate the interaction terms, the corresponding tree-level closed string S-matrix elements in flat Euclidean target space are calculated. First, we observe that $\phi$ is charged under a $U(1)$ symmetry, which corresponds to momentum conservation in the compact direction. Consequently, interaction terms that are charged under this $U(1)$, such as $\phi^2 \phi^*$ cannot appear in the EFT action.

\subsection{The thermal scalar}

This subsection lists the modes that are relevant for the low-energy physics near the Hagedorn temperature. In addition to the massless dilaton, graviton and the antisymmetric tensor,  winding modes with winding numbers $\pm 1$ are light near the Hagedorn temperature. The purpose of this subsection is to review their properties and set up the notations.

String theory at finite temperatures is characterized by a target space with one twisted dimension $X^E$ of circumference
\begin{equation}\label{betaR}
\beta = 2\pi R.
\end{equation}
The radius of the compact dimension is $R$  and the inverse temperature is $\beta$. The closed string spectrum is described by oscillator numbers $N,\widetilde{N}$ associated with the non-compact dimensions, a momentum quantum number $n$ and a winding number $w$ associated with the compact dimension.

The thermal scalar in the closed bosonic string and type II superstring is the state with $N=\widetilde{N}=0,n=0$ and $w=1$. The complex scalar field that corresponds to this state is denoted by $\phi$. Its complex conjugate $\phi^*$ corresponds to  $N=\widetilde{N}=0,n=0$ and $w=-1$. For the heterotic string, $\phi$ and $\phi^*$ correspond to $N=\tilde{N}=0,n=\pm \frac{1}{2}$ and $w=\pm 1$.

The masses-squared of these modes depend on $R$ and the type of string theory \cite{PolchinskiI},\cite{PolchinskiII}:
\begin{eqnarray}
\label{Massphi}
m_{\phi}^2= \left\{
\begin{array}{ll}
 -\frac{4}{\alpha'}+\frac{R^2}{(\alpha') ^2} ~~~\text{bosonic}, \\
-\frac{2}{\alpha'}+\frac{R^2}{(\alpha') ^2} ~~\text{type ~II}, \\
-\frac{3}{\alpha'}+\frac{R^2}{( \alpha')^2}+ \frac{1}{4R^2}
~~~\text{heterotic}.
\end{array} \right.
\end{eqnarray}
The mass vanishes at a particular value of $R$ that corresponds to the Hagedorn temperature $T_H=\beta_H^{-1}$. One can find from equations (\ref{betaR}) and (\ref{Massphi}) that
\begin{equation}
\label{THag}
\beta_H=2\pi \sqrt{\alpha'} \left\{
\begin{array}{ll}
 2 & \text{bosonic} \\
 \sqrt{2} & \text{type ~II} \\
 1+\sqrt{2}/2
 & \text{heterotic}.
\end{array} \right.
\end{equation}
We are interested in temperatures close to $T_H$, so an approximation of the mass squared in this regime can be obtained from Eqs.~(\ref{betaR}-\ref{THag}),
\begin{equation}\label{phimass}
m_{\phi}^2 = -2c_2\epsilon T^2 + O (\epsilon^2),
\end{equation}
where
\begin{equation}
\epsilon= \frac{T-T_H}{T_H}
\end{equation}
and
\begin{equation}\label{c2}
c_2= \left\{
\begin{array}{ll}
64\pi^2 &\text{bosonic} \\
16\pi^2 &\text{type ~II} \\
4\pi^2(4+3\sqrt{2})
&\text{heterotic}.
\end{array} \right.
\end{equation}
According to these equations, the thermal scalar becomes tachyonic for temperatures above the Hagedorn temperature.

The X-CFT part of the vertex operator corresponding to the thermal scalar is the following,
\begin{eqnarray}
\label{vertxTS}
V_{\phi} = g_c ' \int d^2 z :e^{i k_L \cdot X_L+i k_R \cdot X_R} (z,\bar{z}):
\end{eqnarray}
where
\begin{eqnarray}\label{kLkR}
k_L ^E =\frac{R}{\alpha'}~,~
k_R ^E = -\frac{R}{\alpha'}~,~
k_L ^i = k_R ^i ~,~ i=1,...,d,
\end{eqnarray}
where $d$ is the number of non-compact dimensions.
The WS fermion dependence of the operators will be discussed later.
The notations in this equation are: $g_c'$ is a normalization for the vertex operator, $z$ is a WS coordinate, $X_L^{\mu} (z),X_R ^{\mu}(\bar{z})$ are the holomorphic and anti-holomorphic parts of the target space coordinate $X^{\mu} (z,\bar{z})$.  

The normalization of the vertex operator, $g_c '$, is related to the closed string coupling constant $g_c$ by
\begin{equation}
(g_c ') ^2 = g_c ^2 T.
\end{equation}
The string coupling $g_c$ is related to the gravitational coupling $\kappa^2=8\pi G_N$, $G_N$, being Newton's constant  by:
\begin{equation}
\kappa=2\pi g_c
\end{equation}
and the following notation will be useful:
\begin{equation}
\kappa' = 2\pi g_c '.
\end{equation}
To simplify the presentations, we discuss the bosonic string and type II strings from here on. The results for the heterotic string are mentioned when they are relevant.

\subsection{Coupling of two thermal scalars and a dilaton or a graviton}

The purpose of this subsection is to derive the form of the interaction vetrex between a dilaton $\psi$ or a transverse-traceless graviton $h_{\mu \nu}$ and two thermal scalars $\phi,\phi^*$. We show that the corresponding terms in the effective action are
\begin{equation}
S_{int}=-\beta \int d^{d} x ~ 2\kappa'\psi \left(\partial_{\mu} \phi \partial^{\mu}\phi^* + m_{\phi}^2 \phi \phi^* \right),
\end{equation}
 and
\begin{equation}
S_{int}=-\beta \int d^{d} x ~ \kappa' h^{\mu \nu} \partial_{\mu} \phi \partial_{\nu}\phi^*,
\end{equation}
for the dilaton and the graviton, respectively. These terms are deduced  from the tree-level string S-matrix element between two thermal scalars of momenta $k_1,k_2$ and a massless mode of momentum $k_3$ and polarization tensor $\epsilon_{3\mu \nu}$.

A massless mode with vanishing momentum in the compact direction corresponds to the following vertex operator (the superstring vertex is considered in the $(0,0)$ picture):
\begin{eqnarray}
&&V(k,\epsilon)=\frac{2g_c '}{\alpha'} \epsilon_{\mu \nu} \int d^2 z ~\times \cr && \left\{\!\!\!
\begin{array}{ll} ~\partial X^{\mu} \bar{\partial}X^{\nu}e^{ik \cdot X}(z,\bar{z})  ~~~&\text{bosonic} \\
-\left(i\partial X^{\mu} +\frac{1}{2}\alpha' k\cdot \psi \psi^{\mu}\right)\left(i\bar{\partial}X^{\nu} + \frac{1}{2}\alpha' k\cdot \tilde{\psi} \tilde{\psi}^{\nu}\right)e^{ik \cdot X}(z,\bar{z}) ~&\text{type~II}
\end{array}\hspace{.3in} \right.
\end{eqnarray}

The tree-level bosonic S-matrix element is given by:
\begin{eqnarray}\label{psiphiphi*}
\mathcal{S}_3 (k_1,k_2,k_3,\epsilon_3)=(g_c')^2~\frac{2g_c '}{\alpha'} \epsilon_{3\mu \nu} \langle :c\tilde{c} e^{ik_{1L}\cdot X_L + ik_{1R}\cdot X_R}(z_1,\bar{z}_1): \nonumber \\
:c\tilde{c} e^{ik_{2L}\cdot X_L + ik_{2R}\cdot X_R}(z_2,\bar{z}_2):~:c\tilde{c}\partial X^{\mu} \bar{\partial}X^{\nu} e^{ik_3 \cdot X}(z_3,\bar{z}_3):\rangle.
\end{eqnarray}
The notations $c,~\tilde{c}$ represent the ghosts of the closed string. In the type II superstring, the WS fermions terms do not have WS fermion ``partners'' to contract with, hence they do not contribute. Also, there is an additional ghost factor.

We show explicitly the result for the bosonic string. The result for type II strings is the same, as expected. The S-matrix element for the bosonic is string is the following,
\begin{eqnarray}\label{psiphiphi*2}
&\mathcal{S}_3 (k_1,k_2,k_3,\epsilon_3 )=\dfrac{8\pi}{\alpha' (g_c ')^2}~ (g_c ')^2 \dfrac{2g_c '}{\alpha'} 2\pi R (2\pi)^{d}\delta^{d} ({k}_1 ^{\perp} + {k}_2 ^{\perp}+{k}_3 ^{\perp})  \times \nonumber \\
&|z_{12}| ^{2+\alpha' k_{1L} \cdot k_{2L}}|z_{13}| ^{2+\alpha' k_1 \cdot k_3} |z_{23}| ^{2+\alpha' k_2\cdot k_3}\nonumber \\ &\left(-i\dfrac{\alpha'}{2}\right)^2\epsilon_{3 \mu \nu}\left( \frac{k_1 ^{\mu}}{z_{31}}+ \frac{k_2 ^{\mu}}{z_{32}}\right) \left( \dfrac{k_1 ^{\nu}}{\bar{z}_{31}}+ \dfrac{k_2 ^{\nu}}{\bar{z}_{32}}\right).
\end{eqnarray}
Some additional details on the derivation of Eq.~(\ref{psiphiphi*2}) are given in the appendix.
The on-shell conditions and momentum conservation can be used to simplify the expression for $\mathcal{S}_3$. The massless mode satisfies $k_3 ^2 =0$. The scalar products of the momenta are given by
\begin{equation}
k_1 \cdot k_3 = \frac{1}{2}\left((k_1+k_3)^2-k_1 ^2 -k_3 ^2\right)= \frac{1}{2}\left(k_2 ^2 - k_1 ^2 -k_3 ^2\right)= 0,
\end{equation}
\begin{equation}
k_2 \cdot k_3 = \dfrac{1}{2}\left((k_2+k_3)^2-k_2 ^2 -k_3 ^2\right)= \dfrac{1}{2}\left(k_1 ^2 - k_2 ^2 -k_3 ^2\right)= 0,
\end{equation}
\begin{equation}
k_{1L} \cdot k_{2L} = -\frac{R^2}{(\alpha')^2}+{k}_1 ^{\perp} \cdot {k}_2 ^{\perp}.
\end{equation}

Let us consider the limit of small ${k}_1 ^{\perp},{k}_2 ^{\perp}$, relevant to the low-energy EFT, and a compactification scale $R^2 = 4\alpha'$ that corresponds to the bosonic Hagedorn temperature. As mention before, the result for type II superstrings is the same, in spite of the difference in the radius $R^2 = 2\alpha'$, as the picture ghosts compensate for the additional factor.
In this case,
\begin{eqnarray}
\mathcal{S}_3 (k_1,k_2,k_3,\epsilon_3 )=-\beta~4\pi g_c '  (2\pi)^{d}\delta^{d} ({k}_1 ^{\perp} + {k}_2 ^{\perp}+{k}_3 ^{\perp}) \epsilon_{3 \mu \nu} \times \nonumber \\
|z_{12}| ^{-2}|z_{13}| ^{2} |z_{23}| ^{2}\left( \frac{k_1 ^{\mu}}{z_{31}}+ \frac{k_2 ^{\mu}}{z_{32}}\right) \left( \frac{k_1 ^{\nu}}{\bar{z}_{31}}+ \frac{k_2 ^{\nu}}{\bar{z}_{32}}\right).
\end{eqnarray}
Momentum conservation and the fact that the polarization tensor is transverse,
\begin{equation}
\epsilon_{3 \mu \nu} k_3 ^{\mu} = 0 ~,~ k_2 ^{\mu} = -k_1 ^{\mu} -k_3 ^{\mu},
\end{equation}
can be used to simplify the expression for $\mathcal{S}_3(k_1,k_2,k_3,\epsilon_3)$ by eliminating the apparent dependence on the gauge redundant worldsheet coordinates. Recalling that $\kappa' = 2\pi g_c '$, it follows that
\begin{equation}\label{s3TS2D}
\mathcal{S}_3 (k_1,k_2,k_3,\epsilon_3) =-\frac{1}{2}\beta\kappa' (2\pi)^{d}\delta^{d}({k}_1 ^{\perp}+{k}_2 ^{\perp}+{k}_3 ^{\perp}) \epsilon_{3\mu \nu} k_{12} ^{\mu}k_{12} ^{\nu}.
\end{equation}
Here $k_{12} ^{\mu} = k_1 ^{\mu}-k_2 ^{\mu}$. This expression allows us to extract the interaction vertex between a dilaton $\psi$  and two thermal scalars:
\begin{eqnarray}
\label{phiphi*psi}
&S_{int} = \beta \int d^{d} x (-2\kappa'\psi) \left[ \partial_{\mu} \phi \partial^{\mu} \phi^* + m_{\phi} ^2 \phi \phi ^* \right].
\end{eqnarray}
The momenta turn into derivatives and the delta function appears due to the spatial integration.
The mass does not appear in the S-matrix calculation because this was done at the Hagedorn temperature, for which $m_{\phi} ^2=0$. This calculation validates a claim in \cite{HP}, that the interaction between the dilaton and thermal scalar is suppressed by derivatives or by factors of  $\beta-\beta_H$.

For the graviton, $2 \epsilon_{3\mu\nu}$ in (\ref{s3TS2D}) is mapped to $h_{\mu \nu}$ in the effective action. This follows from the graviton's kinetic term, i.e. the second-order Einstein-Hilbert action, which is canonically normalized with that map. The result is thus
\begin{equation}
\label{phiphi*h}
S_{int}=-\beta \int d^{d} x ~ \kappa' h^{\mu \nu} \partial_{\mu} \phi \partial_{\nu}\phi^*.
\end{equation}

\subsection{Four Thermal Scalars}

In this subsection we present a calculation of $(\phi\phi^*)^2$ interaction vertex in the EFT. The result of the calculation is the following quartic term,
\begin{equation}
S_{int}=\beta \int d^{d} x ~ \frac{2\gamma(\kappa')^2}{\alpha'}(\phi \phi^*)^2~,~ \gamma = \left\{
\begin{array}{ll} 1 ~~\text{type II}\\
2 ~~\text{bosonic.} \end{array} \right.
\end{equation}
This interaction term is deduced by comparing to a string S-matrix element calculation in which the exchange of massless modes is subtracted.
The S-matrix calculation begins with the insertion of four vertex operators. The winding numbers are chosen to be $w_1=-w_2=-w_3=w_4=1$. Three of the bosonic coordinates of these operators can be fixed at the following commuting WS coordinates
\begin{equation}\label{gaugefix}
z_2=0~,~z_3=1~,~z_4=w\to \infty.
\end{equation}
This is associated with inserting three pairs of the ghosts $c\tilde{c}$ for each of the fixed coordinates.  The coordinate $z_1$ is integrated.
In the superstring, two of the vertex operators should be taken in the $(-1,-1)$ picture, reflecting the fixing of the anticommuting WS coordinates,  while the rest are in the $(0,0)$ picture.

In equations, the type II vertex operators in the different pictures read:
\begin{eqnarray}\label{VertexOps}
V^{(-1,-1)}=g_c' \int d^2 z~ \delta(\gamma)(z) \delta(\tilde{\gamma})(\bar{z}) e^{ik_L\cdot X_L(z)+ik_R\cdot X_R (\bar{z})}, \nonumber \\
V^{(0,0)} = \frac{\alpha'}{2}g_c' \int d^2 z~ k_L\cdot \psi(z) k_R \cdot \tilde{\psi}(\bar{z})e^{ik_L\cdot X_L(z)+ik_R\cdot X_R (\bar{z})}.
\end{eqnarray}
The $(-1,-1)$ pictures operators have the ghost factors $\delta(\gamma)\delta(\tilde{\gamma})$.
The S-matrix element is thus given by
\begin{eqnarray}\label{Smatrix}
&& \mathcal{S}_4 (\{w_i\}) = \frac{8\pi}{\alpha' (g_c') ^2}(g_c')^4\int_{\mathbb{C}} d^2 z_1 \bigg\langle c\tilde{c}(z_2,\bar{z}_2) c\tilde{c}(z_3,\bar{z}_3)c\tilde{c}(z_4,\bar{z}_4) \nonumber \\
&&   e^{i k_{L1} X_L  (z_1)+ ik_{R1} X_R (\bar{z_1})} e^{i k_{L2} X_L  (z_2)+ ik_{R2} X_R (\bar{z}_2)} e^{i k_{L3} X_L (z_3)+ ik_{R3} X_R (\bar{z_3})} e^{i k_{L4} X_L  (z_4)+ ik_{R4} X_R (\bar{z}_4)} \\
&&    \delta(\gamma)(z_2)\delta(\tilde{\gamma})(\bar{z}_2)\delta(\gamma)(z_3)\delta(\tilde{\gamma})(\bar{z}_3) \frac{(\alpha')^2}{4}k_{1L}\cdot\psi (z_1)k_{R1}\cdot\tilde{\psi} (\bar{z_1})k_{L4}\cdot\psi (z_4)k_{R4}\cdot\tilde{\psi} (\bar{z}_4) \bigg\rangle_{S^2} . \nonumber
\end{eqnarray}
The last factor is special to the superstring. The standard expressions for the correlation functions  \cite{PolchinskiII}, taking into account Eq.~(\ref{gaugefix}), are the following,
\begin{equation}\label{ctildec}
\langle c\tilde{c}(z_2,\bar{z}_2) c\tilde{c}(z_3,\bar{z}_3)c\tilde{c}(z_4,\bar{z}_4)\rangle = |z_{23}|^2|z_{24}|^2|z_{34}|^2=|w|^4,
\end{equation}
\begin{equation}\label{pictures}
\langle \delta(\gamma)(z_2)\delta(\tilde{\gamma})(\bar{z}_2)\delta(\gamma)(z_3)\delta(\tilde{\gamma})(\bar{z}_3)\rangle = 1,
\end{equation}
\begin{equation}
\langle \psi^{\mu} (z_1)\tilde{\psi}^{\nu} (\bar{z}_1) \psi^{\rho} (z_4) \tilde{\psi}^{\sigma} (\bar{z}_4)\rangle = \frac{\delta^{\mu \rho}\delta^{\nu \sigma}}{|z_{14}|^2}=\frac{\delta^{\mu \rho}\delta^{\nu \sigma}}{|w|^2}.
\end{equation}

Next, the X-CFT correlation function is given by
\begin{eqnarray}\label{XCFT}
&\left\langle e^{i k_{L1} X_L  (z_1)+ ik_{R1} X_R(\bar{z_1})}e^{i k_{L2} X_L  (z_2)+ ik_{R2} X_R (\bar{z}_2)} e^{i k_{L3} X_L  (z_3)+ ik_{R3} X_R (\bar{z_3})} e^{i k_{L4} X_L (z_4)+ ik_{R4} X_R (\bar{z}_4)}\right\rangle= \nonumber \\
&\prod_{i<j~i, j=1,...,4}|z_{ij}|^{\alpha' k_{Li} \cdot k_{Lj}}=|w|^{\alpha' k_{L4}\cdot (k_{L1}+k_{L2}+k_{L3})}|z|^{\alpha' k_{L1}\cdot k_{L2}}|1-z|^{\alpha' k_{L1}\cdot k_{L3}}.
\end{eqnarray}
The power of $|w|$ can be found using momentum conservation,
\begin{equation}
|w|^{\alpha' k_{L4}\cdot (k_{L1}+k_{L2}+k_{L3})}=|w|^{-\alpha' k_{L4}^2}=|w|^{-\frac{R^2}{\alpha'}}=\frac{1}{|w|^{2\gamma}}.
\end{equation}
Recall that $R^2 = 2\gamma \alpha'$ is the Hagedorn scale (see Eq.~(\ref{Massphi})). The dependence on $|w|$ cancels out.
Substituting equations (\ref{ctildec}),(\ref{pictures}) and (\ref{XCFT}) into (\ref{Smatrix})
\begin{eqnarray}\label{Smatrix2}
&\mathcal{S}_4 (\{w_i\})= \frac{8\pi(g_c') ^2}{\alpha' }2\pi R (2\pi)^d\delta^{d}\left(\sum_i k^{\perp} _i \right)\times F\times \nonumber \\
&\int_{\mathbb{C}} d^2 z |z|^{\alpha' k_{L1}\cdot k_{L2}}|1-z|^{\alpha' k_{L1}\cdot k_{L3}},
\end{eqnarray}
where
\begin{eqnarray}
F = \left\{
\begin{array}{ll}
1 &\text{bosonic} \\
\left(\frac{\alpha'}{2}k_{L1}\cdot k_{L4}\right)^2 &\text{type ~II}.
\end{array} \right.
\end{eqnarray}

Let us recall the definition of  the Mandelstam variables:
\begin{equation}
s= -(k_1+k_2)^2 ~,~ t= -(k_1+k_3)^2 ~,~ u= -(k_1+k_4)^2,
\end{equation}
and define
\begin{equation}
u^{\perp} = -(k_1 ^{\perp}+k_4 ^{\perp})^2 ~,~ u = -\frac{4R^2}{(\alpha')^2}+u^{\perp}.
\end{equation}
The low-energy limit corresponds to the limit $s,t,u^{\perp}\to 0$.
The relation
\begin{equation}\label{stu}
s+t+u=4m^2
\end{equation}
is satisfied for $m^2 = -{2\gamma}/{\alpha'}$, in which case,
\begin{equation}
s+t+u^{\perp}=4\left(-\frac{2\gamma}{\alpha'}+\frac{R^2}{(\alpha')^2}\right)= 4m_{\phi}^2.
\end{equation}

Now,
\begin{equation}
s=-k_1 ^2 -2k_1 \cdot k_2 -k_2^2 = 2m^2 - 2k_1\cdot k_2= -\frac{4\gamma}{\alpha'}-2k_1\cdot k_2
\end{equation}
implies
\begin{equation}
\alpha' k_{L1}\cdot k_{L2} = -\frac{\alpha'}{2}s-2\gamma.
\end{equation}
Similarly,
\begin{equation}
\alpha' k_{L1}\cdot k_{L3} = -\frac{\alpha'}{2}t-2\gamma,
\end{equation}
and
\begin{equation}
\alpha' k_{L1}\cdot k_{L4} = -\frac{\alpha'}{2}u-2\gamma.
\end{equation}
Using Eq.~(\ref{stu}), it follows that
\begin{equation}
\alpha' k_{L1}\cdot k_{L4} = \frac{\alpha'}{2}(s+t)-2\alpha' m^2-2 = 2+\frac{\alpha'}{2}(s+t).
\end{equation}

We can now express the S-matrix element in terms of $s$, $t$ and $u^{\perp}$,
\begin{eqnarray}\label{Smatrix4}
&\mathcal{S}_4 (\{w_i\})= \frac{8\pi(g_c') ^2}{\alpha' }F\beta (2\pi)^d\delta^{d}\left(\sum_i k^{\perp}_i \right)\int_{\mathbb{C}} d^2 z |z|^{-\frac{\alpha'}{2}s-2\gamma}|1-z|^{-\frac{\alpha'}{2}t-2\gamma} .
\end{eqnarray}
The integral can be evaluated, with the result being
\begin{eqnarray}\label{Smatrix5}
&& \mathcal{S}_4 (\{w_i\})= \frac{16\pi^2(g_c') ^2}{\alpha' }F\beta (2\pi)^d\delta^{d}\left(\sum_i k^{\perp}_i \right)\times
\nonumber \\
&&\frac{\Gamma\left(-\frac{\alpha'}{4}s+1-\gamma\right) \Gamma\left(-\frac{\alpha'}{4}t+1-\gamma\right) \Gamma\left(-1+2\gamma+\frac{\alpha'}{4}s+ \frac{\alpha'}{4}t\right)}{\Gamma\left(2-2\gamma-\frac{\alpha'}{4}s-\frac{\alpha'}{4}t\right) \Gamma\left(\gamma+\frac{\alpha'}{4}s\right)\Gamma\left(\gamma+\frac{\alpha'}{4}t\right)}.
\end{eqnarray}
Next, we use $\kappa'= 2\pi g_c'$, take the limit $s,t\to0$ and obtain
\begin{eqnarray}\label{Smatrix6}
&\mathcal{S}_4 (\{w_i\})= -\frac{4(\kappa') ^2}{\alpha' }\beta (2\pi)^d\delta^{d}\left(\sum_i k^{\perp}_i \right) \nonumber \\ &\left[\frac{\gamma^2}{\frac{\alpha'}{4} s}+\frac{\gamma^2}{\frac{\alpha'}{4} t}+4\gamma+\frac{2\gamma s}{t}+\frac{2\gamma t}{s}+\frac{\alpha' s^2}{4t}+\frac{\alpha' t^2}{4s}+O(s,t)\right].
\end{eqnarray}

In order to find the interaction vertex, the exchange piece should be subtracted. The S-matrix element that results from the exchange of massless modes is the following,
\begin{equation}
\label{Uny}
\mathcal{S}_4 =\frac{1}{2\pi R}\sum_{\epsilon_{\mu\nu}}\int \frac{d^{d} k}{(2\pi)^{d}} \frac{S_3 (k_1,k_2,k,\epsilon_{\mu\nu}) S_{3} (-k,k_3,k_4,\epsilon_{\mu\nu})}{k^2},
\end{equation}
where the three-point amplitude $S_3$ was calculated in Eq.~(\ref{s3TS2D}).

Equation (\ref{Uny}) contains a sum over the polarization tensors. A transverse polarization tensor that satisfies the normalization condition is the following:
\begin{equation}
\epsilon_{\mu \nu} = \frac{k_{12\mu}k_{34\nu}}{\sqrt{k_{12} ^2 k_{34} ^2}}.
\end{equation}
Choosing a basis of tensors that are orthogonal to each other and containing the above element, the sum in (\ref{Uny}) has only one non-vanishing term. Consequently,
\begin{equation}\label{Dilat}
\mathcal{S}_4 =\frac{(\kappa')^2}{4}\beta(2\pi)^{d}\delta^{d}\left(\sum_{i=1} ^4 {k}_i ^{\perp}\right) \frac{ \left((k_1-k_2)\cdot (k_3-k_4)\right)^2}{(k_1+k_2)^2}.
\end{equation}
The denominator corresponds to $s$ and the numerator can be worked out to be $(u-t)^2$. A corresponding $s\to t$ term should also be included, resulting in the following expression,
\begin{eqnarray}\label{Exchange}
&\mathcal{S}_{\text{EXCHANGE}}= -\frac{(\kappa')^2}{4}~\beta (2\pi)^d\delta^{d}\left(\sum_i k^{\perp}_i \right)\left[\frac{(u-t)^2}{s}+\frac{(u-s)^2}{t}\right].
\end{eqnarray}

Using Eq.~(\ref{stu}),
\begin{eqnarray}\label{Exchange2}
&\mathcal{S}_{\text{EXCHANGE}}= -\frac{(\kappa')^2}{4}~\beta (2\pi)^d\delta^{d}\left(\sum_i k^{\perp}_i \right)\left[\frac{\left(\frac{8\gamma}{\alpha'}+s+2t\right)^2}{s}+ \frac{\left(\frac{8\gamma}{\alpha'}+2s+t\right)^2}{t}\right],
\end{eqnarray}
multiplying the numerator and denominator by $\frac{(\alpha')^2}{64}$,
\begin{eqnarray}\label{Exchange3}
&\mathcal{S}_{\text{EXCHANGE}}= -\frac{4(\kappa') ^2}{\alpha'}~\beta (2\pi)^d\delta^{d}\left(\sum_i k^{\perp}_i \right)\left[\frac{\left(\gamma+\frac{\alpha'}{8} s+\frac{\alpha'}{4} t\right)^2}{\frac{\alpha'}{4} s}+\frac{\left(\gamma+\frac{\alpha'}{4} s+\frac{\alpha'}{8} t\right)^2}{\frac{\alpha'}{4} t}\right]
\end{eqnarray}
and expanding in the low-energy limit, we find,
\begin{eqnarray}\label{Exchange4}
&\mathcal{S}_{\text{EXCHANGE}}= -4(\kappa') ^2~\beta (2\pi)^d\delta^{d}\left(\sum_i k_i ^{\perp} \right)\nonumber \\
&\left[\frac{\gamma^2}{\frac{\alpha'}{4} s}+\frac{\gamma^2}{\frac{\alpha'}{4} t}+2\gamma+\frac{2\gamma t}{s}+\frac{2\gamma s}{t}+\frac{\alpha' s^2}{4t}+\frac{\alpha't^2}{4s}+O(s,t)\right].
\end{eqnarray}

The part of $\mathcal{S}_4$ that does not come from the exchange of massless modes can be found by subtracting  Eq.~(\ref{Exchange4}) from Eq.~(\ref{Smatrix6}):
\begin{eqnarray}\label{Smatrix7}
&\mathcal{S}_4 (\{w_i\})= -\frac{8\gamma(\kappa') ^2}{\alpha' }\beta (2\pi)^d\delta^{d}\left(\sum_i k^{\perp}_i \right).
\end{eqnarray}

Finally, the interaction vertex is deduced from taking into account the symmetry factor $4$ and a minus sign due to the minus in $e^{-S_{int}}$ which appears in the path integral. Then, the interaction vertex is given by
\begin{equation}\label{(phi phi*)^2}
S_{int} = \beta\int d^d x \frac{2\gamma(\kappa')^2}{\alpha'} (\phi \phi^*)^2.
\end{equation}
We obtain a positive quartic coupling. This resulted from two minus signs: (a) The difference between the string S-matrix to the string exchange piece turned out to be negative, (b) in the EFT S-matrix $\langle \text{in}|e^{-S_{int}}|\text{out}\rangle$  one expands the exponential to first order $e^{-S_{int}}\sim-S_{int}$.

\section{Effective Action}

Having calculated the interaction vertices, we can now write down the low-energy EFT. The thermal scalar kinetic term,  the interaction with the dilaton and the graviton in Eqs.~(\ref{phiphi*psi}),(\ref{phiphi*h}) together with the quartic self-interaction (\ref{(phi phi*)^2}) can be summarized in the following action,
\begin{eqnarray}
&S_1 = \int d^{d} x \left[\left(\delta^{\mu \nu}-\kappa' h^{\mu \nu}-2\kappa' \delta^{\mu \nu} \psi\right) \partial_{\mu} \phi \partial_{\nu} \phi^* + m_{\phi} ^2 \phi \phi ^* +\frac{2\gamma(\kappa')^2}{\alpha'} (\phi \phi^*)^2\right]
\end{eqnarray}
Redefining $\phi \to \sqrt{\beta}\phi,$ $\kappa'  \psi\to \psi$ and $\kappa'  h_{\mu \nu}\to h_{\mu \nu}$, we find
\begin{eqnarray}
&S_1 = \beta \int d^{d} x\left\{ \left(\delta^{\mu \nu}- h^{\mu \nu}-2\psi\delta^{\mu \nu}\right) \partial_{\mu} \phi \partial_{\nu} \phi^* + m_{\phi} ^2 \phi \phi ^* +\frac{2\gamma\kappa^2}{\alpha'} (\phi \phi^*)^2\right\}.
\end{eqnarray}
Coupling the theory to gravity and the dilaton follows from soft-dilaton result and a scaling consideration \cite{soft-dilaton},  which imply that  $1-2\psi$ should be replaced by  $e^{-2\psi}$. Additionally, the graviton-thermal scalar coupling is replaced by $ \sqrt{G}G^{\mu \nu}\partial_{\mu}\phi \partial_{\nu}\phi^*$,
\begin{equation}
S_1 = \beta\int d^{d} x \sqrt{G}e^{-2\psi}\left(G ^{\mu \nu}\partial_{\mu}\phi \partial_{\nu}\phi^* + m^2 _{\phi}\phi \phi^*+\frac{2\gamma \kappa^2}{\alpha'}(\phi \phi^*)^2\right).
\end{equation}

It is convenient to rewrite the mass using Eq.~(\ref{phimass}) and the quartic-self interaction as:
\begin{eqnarray}
&m_{\phi}^2 = -2c_2\epsilon T^2+O(\epsilon^2),\\
&\frac{2\gamma\kappa^2}{\alpha'} = c_2 \kappa^2 T^2+O(\epsilon).
\end{eqnarray}
The numerical constant $c_2$ can be found in Eq. (\ref{c2}).

Then, to leading power in $\epsilon$,
\begin{eqnarray}
&S_1 = \beta\int d^{d} x \sqrt{G}e^{-2\psi}\left\{G ^{\mu \nu}\partial_{\mu}\phi \partial_{\nu}\phi^* - 2c_2 \epsilon T^2\phi \phi^*+c_2 \kappa^2 T^2(\phi \phi^*)^2\right\}.
\end{eqnarray}

We can add to the EFT action for the thermal scalar the standard action for the graviton and dilaton,
\begin{eqnarray}
&S_2 = -\frac{\beta}{2\kappa^2} \int d^{d} x \sqrt{G} e^{-2\psi} \biggl\{R-2\Lambda+4G^{\mu \nu}\partial_{\mu} \psi \partial_{\nu} \psi\biggr\},
\end{eqnarray}
where $\Lambda$ is the cosmological constant (CC). The CC needs to be added so that the EFT S-matrix elements match the corresponding string S-matrix elements. The value of the CC is fixed in the next subsection and its addition is further justified in Section 4.
Terms involving the antisymmetric tensor and the graviphoton could have been added to the action, but we set them to zero in the current discussion. The metric compact-compact component is fixed by the thermal environment to unity and does not fluctuate.

The total EFT action is therefore given the sum of $S_1$ and $S_2$,
\begin{eqnarray}
S &=& \beta\int d^{d} x \sqrt{G}e^{-2\psi}\Biggl\{-\frac{1}{2 \kappa^2}R+\frac{1}{\kappa^2}\Lambda- \frac{2}{\kappa^2}G^{\mu \nu}\partial_{\mu}\psi\partial_{\nu} \psi  \cr &+&  G ^{\mu \nu}\partial_{\mu}\phi \partial_{\nu}\phi^* - 2c_2 \epsilon T^2\phi \phi^*+c_2 \kappa^2 T^2(\phi \phi^*)^2\Biggr\}.
\label{totalEFT}
\end{eqnarray}
The thermal scalar part of the action, assuming a flat metric and constant dilaton, was found in \cite{CFT4dS} in a parametric form. Our results verify the qualitative form of the action and fix the coefficients in terms of $\kappa^2$ and $\epsilon$.

\subsection{Equations of motion and their solution}

The equation of motion (EOM) for $\phi^*$, multiplied by $\phi^*$, is the following,
\begin{equation}
\phi^*\nabla^2 \phi =-2c_2 \epsilon T^2 \phi \phi^* + 2c_2 \kappa^2 T^2 (\phi \phi^*)^2.
\end{equation}
For $\epsilon>0$ a stable solution is $\phi \phi^*(x) = \dfrac{\epsilon}{\kappa^2}$, while for $\epsilon<0$ the solution is $\phi=0$ . Therefore,
\begin{equation}\label{phi0}
\phi \phi^*(x) = \frac{\epsilon}{\kappa^2} \Theta(\epsilon),
\end{equation}
where $\Theta$ is the Heaviside function. Next, the dilaton EOM is
\begin{eqnarray}
\label{dilaton}
&R-2\Lambda+4 \nabla^2\psi -4\partial^{\mu} \psi \partial_{\mu} \psi = \nonumber \\
&  2\kappa^2 \left[ \partial^{\mu} \phi \partial_{\mu} \phi^* -2c_2\epsilon T^2\phi \phi^*+c_2\kappa^2 T^2(\phi \phi^*)^2\right]
\end{eqnarray}
and the metric EOM is
\begin{eqnarray}
\label{metricEOM}
&R_{\mu \nu}-\frac{1}{2} G_{\mu\nu}(R-2\Lambda)+4\partial_{\mu} \psi \partial_{\nu} \psi-2G_{\mu \nu}  \partial^{\alpha} \psi \partial_{\alpha} \psi =\nonumber \\
& 2\kappa^2\left(\partial_{\mu} \phi \partial_{\nu} \phi^*-\frac{1}{2}  G_{\mu \nu}\left[\partial^{\mu} \phi \partial_{\mu} \phi^* -2\epsilon c_2 T^2 \phi \phi^*+c_2\kappa^2 T^2(\phi \phi^*)^2\right]\right).
\end{eqnarray}
The dilaton EOM can be used to simplify Eq.~(\ref{metricEOM}),
\begin{eqnarray}
&R_{\mu \nu}+4\partial_{\mu} \psi \partial_{\nu} \psi+2G_{\mu \nu}\nabla^2 \psi-4G_{\mu \nu}\partial^{\alpha} \psi \partial_{\alpha} \psi=
&2\kappa^2~\partial_{\mu} \phi \partial_{\nu} \phi^*.
\end{eqnarray}

Focusing on $\epsilon>0$, a background solution of the EOM is given below,
\begin{eqnarray}
\label{background}
\phi\phi^*(x)&=&\dfrac{\epsilon}{\kappa^2}, \nonumber \\
G_{\mu \nu}(x)&=&\delta_{\mu \nu},  \\
\psi(x)&=&\text{const}. \nonumber
\end{eqnarray}
The value of the CC is given by
\begin{equation}
\Lambda=c_2\epsilon^2 T^2.
\end{equation}

The value of the CC is chosen to ensure a d-dimensional flat space solution, such that the EFT background corresponds to the string theory flat background. The consistency of the background solution is further validated by constructing a corresponding WS CFT.  The constant dilaton is chosen to ensure a finite but small string coupling.

We can expand the action $S_1$ about the solution for $T>T_H$,
\begin{equation}
\phi = \frac{\sqrt{\epsilon}}{\kappa}+ \varphi
\end{equation}
and find the EFT potential relevant to perturbation theory about the background (\ref{background}),
\begin{eqnarray}
\label{VarphiPotential1}
V = -\dfrac{c_2 \epsilon^2 T^2}{\kappa^2} + 2c_2 \epsilon T^2 \varphi \varphi^* +
2\sqrt{\epsilon} c_2 \kappa T^2 \varphi \varphi^* (\varphi +\varphi^*)+c_2\kappa^2 T^2 (\varphi \varphi^*)^2.
\end{eqnarray}
Here, cubic terms are allowed because the $U(1)$ symmetry is spontaneously broken. As $\phi$ condenses, the original $U(1)$ symmetry is broken to a discrete $Z_2$ symmetry, $\varphi\leftrightarrow\varphi^*$.

In the next Section we write down a worldsheet theory corresponding to the background in Eq.~(\ref{background}).

\subsection{Massless momentum mode of the bosonic string}
In this subsection, we consider the bosonic string and ignore momentarily the closed string tachyon, focusing on the dynamics of massless modes near the Hagedorn temperature.
In contrast to the type II superstring, the bosonic string has momentum modes which become massless at the Hagedorn temperature. These are the fourth momentum modes of the tachyon, which we label as $T_4$ and $T_4^*$.  Their mass squared is given by,
\begin{equation}
\label{MassT4}
m_{T_4} ^2 = -\frac{4}{\alpha'}+\frac{16}{R^2}.
\end{equation}
Using $T_H^{-1}=4\pi \sqrt{\alpha'}$ and $\beta = 2\pi R$, it follows that
\begin{equation}
m_{T_4} ^2 = 64\pi^2 (T^2-T_H^2).
\end{equation}
Consequently, these momentum modes are massless at the Hagedorn temperature. For $T\approx T_H$, the mass-squared is given by
\begin{eqnarray}
\label{T4mass}
m_{T_4}^2 &=& 2c_2 \epsilon T^2 + O (\epsilon^2), \cr
 c_2&=&64\pi^2.
\end{eqnarray}
The momentum modes $T_4$, have the same self-interaction as that of the thermal scalar,
\begin{equation}
S_{int}=\beta \int d^{d} x ~ \frac{4(\kappa')^2}{\alpha'}(T_4 T_4^*)^2.
\end{equation}

It is also possible to find the interaction vertex of $\phi\phi^* T_4 T_4 ^*$. The calculation follows almost exactly the calculation of the self-interaction, the difference being that the symmetry factor of $4$ is absent because all the particles involved are different. The result is then
\begin{equation}
S_{int} = \beta \int d^{d}x \frac{16(\kappa')^2}{\alpha'}\phi \phi^* T_4 T_4 ^*.
\end{equation}

We can combine the $T_4$ vertices and the previously obtained thermal scalar vertices into a single effective action,
\begin{eqnarray}
&S_{\phi,T_4} = \beta\int d^{d} x \sqrt{G}e^{-2\psi}\Biggl\{G ^{\mu \nu}\partial_{\mu}\phi \partial_{\nu}\phi^* +m_{\phi}^2\phi \phi^*+c_2 \kappa^2 T^2(\phi \phi^*)^2+ \cr
& G ^{\mu \nu}\partial_{\mu}T_4 \partial_{\nu}T_4^* + m_{T_4}^2 T_4 T_4^*+c_2 \kappa^2 T^2(T_4 T_4^*)^2+4c_2 \kappa^2 T^2 \phi \phi^* T_4 T_4 ^*\Biggr\}.
\label{t4phiact}
\end{eqnarray}
This action has a $U(1)\times U(1)\times Z_2$ internal symmetry, which is explicitly broken down from a $U(2)$ symmetry by the last term.

The EOM resulting from the action in Eq.~(\ref{t4phiact}) combined with the metric and dilaton EOMs, yield the following background solution,
\begin{eqnarray}
\label{backgroundT4}
\phi\phi^*(x)&=&\frac{\epsilon}{\kappa^2}\Theta(\epsilon)\cr
T_4 T_4^*(x)&=&-\frac{\epsilon}{\kappa^2}\Theta(-\epsilon), \nonumber \\
G_{\mu \nu}(x)&=&\delta_{\mu \nu} \cr
\psi(x)&=&\text{const},
\end{eqnarray}
where value of the CC is given by
\begin{equation}
\Lambda=c_2\epsilon^2 T^2.
\end{equation}
As either $\phi$ or $T_4$ condense, the $U(1)\times U(1)\times Z_2$ symmetry is broken to the product of $U(1)$ of the field that does not condense and a $Z_2$ of the condensing field.

As for the case of the thermal scalar, this background can be validated using a WS CFT, as explained in the next section.

\section{Worldsheet conformal field theory}
As a solution of tree-level string theory, the background written above must correspond to a worldsheet (WS) 2D CFT. We wish to show that this is indeed the case.

Let us discuss the following 2D action,
\begin{equation}
\label{WS2D}
S = S_0 +S_{\text{ghosts}}+\int d^2 z~ \mu \left(O_{1} (z,\bar{z})+O_{-1}(z,\bar{z})\right),
\end{equation}
where
\begin{equation}
S_0=\frac{1}{4\pi} \int d^2 z \left[\frac{2}{\alpha'}\partial X^{\mu} \bar{\partial} X_{\mu}+\psi^{\mu}\bar{\partial}\psi_{\mu} +\tilde{\psi}^{\mu}\partial\tilde{\psi}_{\mu}
\right],
\end{equation}
$S_{\text{ghosts}}$ denotes the ghost CFT action and the operators $O_{\pm1}(z,\bar{z})$ are the vertex operators of the $\pm 1$ winding modes defined in Eqs.~(\ref{kLkR}),(\ref{VertexOps}),
\begin{equation}
O_{\pm 1} = \frac{\alpha'}{2} k_L\cdot \psi(z) k_R \cdot \tilde{\psi}(\bar{z})e^{ik_L\cdot X_L(z)+ik_R\cdot X_R (\bar{z})}.
\end{equation}
The action in Eq.~(\ref{WS2D}) can be therefore seen to be given by,
\begin{equation}
\label{WS2DCos}
S = S_0 +S_{\text{ghosts}}+\int d^2 z~ 2 \mu ~\frac{\alpha'}{2} e^{ik^{\perp}\cdot X(z,\bar{z})}   k_L\cdot \psi(z) k_R \cdot \tilde{\psi}(\bar{z}) \cos\left[\frac{R}{\alpha'}\left(X_L^E-X_R^E\right) \right].
\end{equation}
Worldsheet fermions factors are absent in the bosonic string. Picture changing operators can be inserted seprately when calculating correlation functions.
Maggiore \cite{Maggiore} used the action (\ref{WS2DCos}) in order to describe the physics above $T_H$, that he argued includes a quantization of the temperature as well as non-commutative spatial geometry.

The parameter $\mu$ in Eq.~(\ref{WS2DCos}) is determined as we now explain.
The S-matrix elements calculated in the background (\ref{background}) need to match the WS CFT correlation functions. Below we show that for a particular value of $\mu$, $\mu=\sqrt{\epsilon}$, the 1-,2-,3- and 4-point functions of the EFT  agree with the correlation function of the WS theory in Eq.~(\ref{WS2D}). The agreement is verified  to leading order in the expansion that we already used when deriving the EFT action, namely, in $\epsilon$ and in the dimensionless string coupling. In this case the parameters are small and conformal perturbation theory is applicable. The operators $O_{\pm1}$ are marginal operators of the Polyakov action and so, in conformal perturbation theory, the modified action in Eq.~(\ref{WS2D}) is also a CFT.

Since the string EFT is only accurate to leading order in the above expansion, we cannot expect an exact 2D CFT to correspond to the EFT, so matching in conformal perturbation theory is the best that we can get.

\subsection{Two-point function}

First, we discuss the matching of the WS CFT and EFT two-point functions.  On the EFT side, from Eq.~(\ref{VarphiPotential1}), we can deduce that the two-point function is the following,
\begin{equation}
G_{\varphi\varphi^*}(k_1,k_2) = 2c_2\epsilon T^2 \cdot \beta  (2\pi)^d\delta^{d}\left(k_1^{\perp}+k_2^{\perp}\right).
\end{equation}
The zeroth order two-point function $G_{\varphi\varphi^*}^{[0]}(k_1,k_2)$, is that of the free CFT,
\begin{equation}
G_{\varphi\varphi^*}^{[0]}(k_1,k_2) =\int d^2 z_1 d^2 z_2 \langle O_{1}(k_1,z_1,\bar{z}_1) O_{-1}(k_2,z_2,\bar{z}_2)\rangle,
\end{equation}
evaluated with the action $S_P$,
\begin{equation}
G_{\varphi\varphi^*}^{[0]}(k_1,k_2) = -2c_2\epsilon T^2 \cdot \beta (2\pi)^d \delta^{d}\left(k_1^{\perp}+k_2^{\perp}\right),
\end{equation}
which has the apparent tachyonic behavior above $T_H$. The calculation below is valid for both the bosonic and the superstring, that differ only by the values of $c_2$.

To first order in $\mu$, the contribution to $G_{\varphi\varphi^*}(k_1,k_2)$ vanishes, because the winding number is not conserved. To second order in $\mu$, two terms satisfy winding number conservation:
\begin{eqnarray}
&G_{\varphi\varphi^*}^{[2]}(k_1,k_2) =\dfrac{\mu^2}{2}\int d^2 z_1...d^2 z_4 \left\langle O_{1}(k_1,z_1,\bar{z}_1) O_{-1}(k_2,z_2,\bar{z}_2)\times \right. \nonumber \\
&\left.\bigl[O_{1} (k_3,z_3,\bar{z}_3) O_{-1}(k_4,z_4,\bar{z}_4)+O_{-1} (k_3,z_3,\bar{z}_3)O_{1}(k_4,z_4,\bar{z}_4)\bigr]\right\rangle.
\end{eqnarray}
The two terms are equal to each other and were computed for the two-to-two S-matrix of four thermal scalars in Sect.~2.3. After subtracting a divergent exchange term, we obtain,
\begin{equation}
G_{\varphi\varphi^*}^{[2]}(k_1,k_2) =4c_2\mu^2 T^2 \cdot \beta (2\pi)^d \delta^d (k_1^{\perp}+k_2^{\perp}).
\end{equation}
Thus, setting
\begin{equation}\label{mu2}
\mu^2 = \epsilon,
\end{equation}
the WS two-point function to second order is given by
\begin{equation}
G_{\varphi\varphi^*}(k_1,k_2) =2c_2\epsilon T^2 \cdot \beta (2\pi)^d \delta^d \left(k_1^{\perp}+k_2^{\perp}\right).
\end{equation}
This agrees with the EFT two-point function in Eq.~(\ref{VarphiPotential1}).

\subsection{1,3 and 4-point functions}
\textbf{One-point function}

To zeroth order (and to any even order), because of winding number conservation,
\begin{equation}
\left\langle \int d^2 z_1 ~O_1(k_1,z_1,\bar{z}_1) \right\rangle ^{[0]}=0.
\end{equation}
To first order,
\begin{eqnarray}
\label{1stonep}
&\left\langle \int d^2 z_1 O_1(k_1,z_1,\bar{z}_1) \right\rangle ^{[1]}=\int d^2 z_1 \int d^2 z_2 ~\mu \left\langle O_1(k_1,z_1,\bar{z}_1) O_{-1}(k_2,z_2,\bar{z}_2)\right\rangle =\nonumber \\
&-2c_2 \mu\epsilon T^2 \cdot \beta (2\pi)^d \delta^d \left(k_1^{\perp}+k_2^{\perp}\right),
\end{eqnarray}
and to third order,
\begin{eqnarray}
\label{3rdonep}
&\left\langle \int d^2 z_1 O_1(k_1,z_1,\bar{z}_1) \right\rangle ^{[3]}=3\int d^2 z_1... \int d^2 z_4 ~\frac{\mu^3}{3!}\times \nonumber \\
&\left\langle O_1(k_1,z_1,\bar{z}_1) O_{-1}(k_2,z_2,\bar{z}_2)O_1(k_3,z_3,\bar{z}_3)O_{-1}(k_4,z_4,\bar{z}_4)\right\rangle =\nonumber \\
&2 c_2 \mu^3   T^2 \cdot \beta (2\pi)^d \delta^d \left(k_1^{\perp}\right).
\end{eqnarray}
The contractions need to be done such that they conserve winding number.

Combining Eq.~(\ref{1stonep}) and (\ref{3rdonep}) we conclude that up to 3rd order, the one-point function vanishes for $\mu=\sqrt{\epsilon}$. The 4th order correction vanishes and higher orders are suppressed by powers of $\epsilon$, so they do not affect the comparison since their coefficients can be affected by higher order corrections to the EFT potential, which we have not taken into account.

\textbf{Three-point function}

The zero'th order contribution to the CFT three-point function vanishes, due to winding number conservation,
\begin{equation}
\left\langle \int d^2 z_1 \int d^2 z_2 \int d^2 z_3 ~O_1(k_1,z_1,\bar{z}_1) O_{-1}(k_2,z_2,\bar{z}_2)O_1(k_3,z_3,\bar{z}_3)\right\rangle ^{[0]}=0.
\end{equation}

The first order contribution is the following,
\begin{eqnarray}
&&\left\langle \int d^2 z_1 \int d^2 z_2 \int d^2 z_3 ~O_1(k_1,z_1,\bar{z}_1) O_{-1}(k_2,z_2,\bar{z}_2)O_1(k_3,z_3,\bar{z}_3)\right\rangle ^{[1]}\nonumber \\
&=&\mu \left\langle \int d^2 z_1...d^2 z_4 ~O_1(k_1,z_1,\bar{z}_1) O_{-1}(k_2,z_2,\bar{z}_2)O_1(k_3,z_3,\bar{z}_3)O_{-1} (k_4,z_4,\bar{z}_4)\right\rangle  \nonumber \\
&=&  4\kappa  \mu c_2  T^2 \cdot \beta (2\pi)^d \delta^d \left(\sum_i k_i^{\perp}\right).
\end{eqnarray}
Substituting $\mu = \sqrt{\epsilon}$ we find
\begin{equation}
4 \kappa \mu c_2  T^2 =4 \sqrt{\epsilon} c_2 \kappa T^2.
\end{equation}
Comparing to the EFT in Eq.~(\ref{VarphiPotential1}) and recalling that there is a symmetry factor of 2 when evaluating the EFT correlation functions, we find agreement with Eq.~(\ref{mu2}) ~ \footnote{The reason for the $\kappa$ is that the S-matrix calculation of 4 winding modes involves a factor $\kappa^2$, but the fourth insertion from the perturbation theory lacks a normalization with $\kappa$, ultimately giving rise to $\kappa^1$. }.

\textbf{Four-point function}

The leading order correction to the four-point function is given by
\begin{eqnarray}
&\left\langle \int d^2 z_1...d^2 z_4 ~O_1(k_1,z_1,\bar{z}_1) O_{-1}(k_2,z_2,\bar{z}_2)O_1(k_3,z_3,\bar{z}_3)O_{-1} (k_4,z_4,\bar{z}_4)\right\rangle^{[0]} \nonumber \\
&= 4c_2 \kappa^2  T^2 \cdot \beta (2\pi)^d \delta^d \left(\sum_i k_i^{\perp}\right),
\end{eqnarray}
where the factor of $4$ is the symmetry factor. The result is consistent with the EFT S-matrix element upon substituting $\mu = \sqrt{\epsilon}$.

\section{Generalized conformal structure of the target space EFT}

We wish to show that the EFT is conformally invariant when the temperature varies in a specific way under the conformal transformation. Typically, field theories at finite temperatures are not conformal, because the temperature is not considered to transform under Weyl transformations. However, the temperature in our setup depends on the compact-compact component of the metric, so it does transform under Weyl transformations.  This feature of the EFT is referred to as a ``generalized conformal structure'', see \cite{Jevicki1},\cite{Jevicki2},\cite{Skenderis}.

The invariance of the EFT is significant because it guarantees that the correlation functions of the EFT obey the Ward identities of the conformal symmetry and in particular fixes the two- and three-point functions up to constants.

We start by showing the EFT is scale invariant and then we show that a simple change of variables makes it, to the lowest order in the expansion parameters, also conformally invariant.

\subsection{Scaling Symmetry}

The purpose of this subsection is to show that the effective theory is scale invariant when the temperature transforms appropriately.

Recall the effective action of the thermal scalar:
\begin{equation}\label{Eaction}
S = \beta\int d^{d} x \sqrt{G}e^{-2\psi}\left(G ^{\mu \nu}\partial_{\mu}\phi \partial_{\nu}\phi^* -2c_2\epsilon T^2 \phi \phi^*+c_2\kappa^2 T^2(\phi \phi^*)^2\right).
\end{equation}
The square root of the metric $\sqrt{G}$ includes the factor $\sqrt{G_{\tau \tau}}$ where $\tau$ parametrizes the thermal circle. The inverse temperature, is by definition related to $\sqrt{G_{\tau \tau}}$ as follows,
\begin{equation}
\frac{1}{T} \equiv \int_0 ^{\beta} \sqrt{G_{\tau \tau}} d\tau=\beta \sqrt{G_{\tau \tau}}.
\end{equation}
The last equality holds for $\tau$-independent metrics.

Let us consider a Weyl transformation, under which the metric and the thermal scalar transform as:
\begin{eqnarray}
G_{\mu \nu} &\to& \Omega^2 G_{\mu \nu}, \cr
\phi &\to& \Omega^{-\frac{d-1}{2}}\phi,
\end{eqnarray}
for a real, space-independent $\Omega$. The temperature transforms as:
\begin{equation}
\frac{1}{T} = \beta \sqrt{G_{\tau \tau}}\to \Omega \frac{1}{T}.
\end{equation}

To demonstrate that the action in Eq.~(\ref{Eaction}) is invariant under scaling transformations, we consider the mass  and the quartic interaction terms. It is easy to check that the kinetic term is scale invariant.
For the mass term it is important to note that $\epsilon$ stays invariant under a Weyl transformation because its numerator and denominator transform the same way. Then,
\begin{equation}
\int d^{d} x \sqrt{G}e^{-2\psi}  \epsilon T^2 \phi \phi^*\to \Omega^{d+1-2-(d-1)}\int d^{d} x \sqrt{G}e^{-2\psi}\epsilon T^2 \phi \phi^*,
\end{equation}
hence the mass term is also invariant.  Next, we are interested on the quartic term. It is first useful to consider the transformation of $\kappa$. Keeping the dimensionless string coupling constant $g_s$ fixed under Weyl transformations,
\begin{equation}
\kappa^2 \propto g_s ^2 \left(\alpha'\right)^{\frac{d-1}{2}} \propto \frac{g_s^2}{T ^{d-1}}\to \Omega^{d-1} \kappa^2.
\end{equation}
Therefore, the interaction term transforms as,
\begin{eqnarray}
&\int d^{d} x \sqrt{G}e^{-2\psi} \kappa^2 T^2(\phi \phi^*)^2\to\nonumber \\ &\Omega^{d+1+d-1-2-2(d-1)}\int d^{d} x \sqrt{G}e^{-2\psi} \kappa^2 T^2(\phi \phi^*)^2.
\end{eqnarray}
Thus, the quartic interaction term is also invariant.
We conclude that the EFT action is scale invariant.
The same calculations apply if we include the momentum modes in the bosonic string.

\subsection{Conformal Symmetry}

Let us now consider a conformal transformation with a space-dependent Weyl factor $\Omega=\Omega(x)$. Then the kinetic term in eq.~(\ref{Eaction}) is not invariant, while the potential terms remain invariant. A conformal coupling of the form
\begin{equation}
S_{conf}=\xi \beta \int d^d x \sqrt{G} e^{-2\psi} \mathcal{R} \phi \phi^* ~,~ \xi = \frac{d-2}{4(d-1)},
\end{equation}
where $d$ is the number of non-compact spatial dimensions, would render the action Weyl-invariant \cite{BD}.

While the string frame has no conformal coupling term, the latter can be introduced by a metric redefinition, which does not alter solutions to the EOMs and S-matrix elements to leading order in  $\kappa^2$,
\begin{equation}
G_{\mu \nu} = e^{2\omega(x)} \widehat{G}_{\mu \nu},
\end{equation}
\begin{equation}
e^{2\omega(x)}= 1+a_0 \kappa^2 \phi \phi^*(x).
\end{equation}
The string frame metric is $G_{\mu \nu}$ whereas the ``conformal frame'' metric is  $\widehat{G}_{\mu \nu}$.  The parameter $a_0$ is a $d$-dependent constant which will be chosen to reproduce the conformal coupling, as will be discussed below.
Recall that we are working to lowest order in the expansion, so ${2\omega(x)}\simeq a_0 \kappa^2 \phi \phi^*(x)$.

This redefinition relates the determinants of the metrics,
\begin{equation}
\sqrt{G} = \left[1+a_0 \kappa^2 \left(\phi \phi^*(x)\right)\right]^\frac{d}{2} \sqrt{\widehat{G}},
\end{equation}
which, to leading order in $\kappa^2$ is given by,
\begin{equation}
\sqrt{G} = \left[1+\frac{d a_0}{2}\kappa^2  \phi \phi^*(x) \right] \sqrt{\widehat{G}}.
\end{equation}
The Ricci scalar transforms as
\begin{equation}\label{FRAMES}
\mathcal{R}=e^{-2\omega}\left(\widehat{\mathcal{R}}-2(d-1)\nabla^2 \omega - (d-2)(d-1)\partial_{\mu} \omega \partial^{\mu} \omega\right).
\end{equation}
The last term in this equation is $O(\kappa^4)$, beyond the order $O(\kappa^2)$ that we are interested on. It is further suppressed by derivatives. Thus, we can neglect it. The second term transforms as
\begin{equation}
\nabla^2 \omega=\frac{a_0 \kappa^2}{2} \left(2\partial_{\mu}\phi \partial^{\mu} \phi^*(x)+\phi \nabla^2 \phi^* +\phi^* \nabla^2 \phi\right)+O(\kappa^4).
\end{equation}
We observe an order $\kappa^2$ correction to the thermal scalar kinetic term, which can be neglected. The other two terms to order $\kappa^2$ can be cancelled by subtracting these terms from the action, this freedom is justified because the S-matrix elements would not change, as $\nabla^2 \phi\sim k^2 \phi_k=0$ on-shell.

Expanding Eq.~(\ref{FRAMES}), we find,
\begin{eqnarray}
&\mathcal{R} = \widehat{\mathcal{R}}-a_0 \kappa^2 \widehat{\mathcal{R}}\phi \phi^*(x)+\cdots.
\end{eqnarray}
As a result, the ``Einstein-Hilbert'' term in the string frame becomes, to leading order,
\begin{eqnarray}
&-\frac{\beta}{2\kappa^2} \int d^d x \sqrt{G} e^{-2\psi} \mathcal{R} = -\frac{\beta}{2\kappa^2} \int d^d x \sqrt{\widehat{G}} e^{-2\psi} \widehat{\mathcal{R}}\nonumber \\
&-\frac{a_0}{2}\beta \left(\frac{d}{2}-1\right)\int d^d x \sqrt{\widehat{G}} e^{-2\psi} \widehat{\mathcal{R}} \phi \phi^*.
\end{eqnarray}
In order to have conformal coupling, one demands
\begin{equation}
-\frac{a_0}{2}\left(\frac{d}{2}-1\right)=\frac{d-2}{4(d-1)}.
\end{equation}
Therefore,
\begin{equation}
a_0 = -\frac{1}{d-1}.
\end{equation}
The rest of the terms in the action are not affected to leading order. The mass term, being multiplied by $\sqrt{G}$, gets a correction of order $\sim \epsilon \kappa^2$, i.e. a subleading contribution. Similarly, the quartic interaction term is unchanged to leading order.

The conclusion is that to leading order in the small parameters, the hatted-frame exhibits conformal symmetry, while still having the same solutions to EOMs and the same S-matrix elements as those we obtained previously.

\section{The Hagedorn phase transition}

This section is about the nature of the transition for the bosonic string and the type II superstring. Here, we are interested in the dependence of the string free energy at temperatures just below the Hagedorn transition and just above the Hagedorn temperatures, keeping the temperature fixed in each of the evaluations, thus, the thermal circle has a fixed circumference.

The Euclidean action is commonly identified with the Helmholtz free energy $F$ and the Euclidean Lagrangian is therefore identified with the Helmholtz free energy density. However, in the case that we consider, we also have a CC term in the action $\sim \Lambda \int d^D x \sqrt{g}$ which can be viewed as ``pressure times volume'' term in the thermodynamic potential \cite{CCpressure1,CCpressure2,CCpressure3}.  The addition of the CC can therefore be viewed as a Legendre transform of the Helmholtz free energy, transforming it into the Gibbs free energy. Consequently, the Lagrangian density of the thermal scalar  corresponds to the Helmholtz free energy density while the full Lagrangian density which includes the CC should be identified with the Gibbs free energy density. To determine the nature of the transition we can therefore either include the CC and use the Gibbs free energy or exclude it and use the Helmholtz free energy. In what follows use the thermal scalar action to calculate the Helmholtz free energy.

Previously, it was argued that if the size of the thermal circle can fluctuate, then the transition is first order \cite{AW},\cite{Gubser},\cite{Barbon1}.  We start by reviewing this result.

Our calculation in Section 2.2 which led to Eq.~(\ref{s3TS2D}) of the S-matrix element of two thermal scalar and one massless mode, implies an interaction vertex of the radion $h_{\tau \tau}$ and two thermal scalars:
\begin{equation}\label{htautau}
S_{int} = \beta \int d^dx \frac{\beta^2}{4\pi^2 (\alpha')^2}h_{\tau \tau} \phi \phi^*.
\end{equation}
To see this, substitute $k_L ^E = \frac{R}{\alpha'}=\frac{\beta}{2\pi \alpha'}$ and $2e_{\tau \tau} \to h_{\tau \tau}$ in Eq.~(\ref{s3TS2D}).
Varying Eq.~(\ref{htautau}) and the Einstein-Hilbert term with respect to $h_{\tau \tau}$, one obtains the following EOM,
\begin{equation}
\nabla^2 h_{\tau \tau} (x)= \frac{\beta^2}{\pi^2 \alpha'}\kappa^2 \phi \phi^* (x).
\end{equation}
This equation can be solved using the Green's function of the Laplace operator $G_{\nabla^2}(x,x')$ and $h_{\tau \tau}$ can be integrated out of Eq.~(\ref{htautau}). The result is a non-local, negative quartic self-interaction for the thermal scalar:
\begin{equation}
V(\phi,\phi^*) (x) = -\frac{\beta^4}{4\pi^4 (\alpha')^4}\kappa^2\phi \phi^*(x)\int G_{\nabla^2}(x,x')\phi \phi^* (x') d^d x'.
\end{equation}
In this case, the phase transition is first order because at $T=T_H$, the effective potential, albeit being non-local, is decreasing at the origin of the $\phi,\phi^*$ field space. Higher order terms in the effective potential (like $(\phi \phi^*)^6,(\phi \phi^*)^8...$) are expected to stabilize it, giving rise to minima of the potential. Transitions between $T<T_H$ to $T_H<T$ requires latent heat, the defining feature of first-order phase transitions. 

Next, consider a system of bosonic closed strings in thermal equilibrium, thus the radion is set to zero by a Lagrange multiplier. The transition is a continuous crossover as the free energy density $F(T)$ behaves the same way in both $T<T_H$ and $T>T_H$, at least for the background we found in Eq.~(\ref{background}):
\begin{equation}
F=\mathcal{L}|_{\text{background}}~,~ F(T_H(1+\epsilon))=F(T_H(1-\epsilon)).
\end{equation}
Here, $\mathcal{L}$ is the Lagrangian density. The last equation is a result of the momentum modes $T_4$ and $T_4 ^*$ that play the role of the winding modes below the Hagedorn temperature.

In contrast, type II superstrings in thermal equilibrium undergo a second-order phase transition. This can be seen using the Ehernfest criterion:
\begin{equation}
\frac{d^2 F}{dT^2} |_{T_H ^-} \neq \frac{d^2 F}{dT^2} |_{T_H ^+},
\end{equation}
while $F$ and its first derivatives are continuous.

Below the Hagedorn, $T_H^-$, the associated background $G_{\mu \nu}=\delta_{\mu \nu},\psi=\text{const},\phi=\phi^*=0$ implies that
\begin{equation}
F(T_H^-)=\mathcal{L}|_{\text{background}}=0.
\end{equation}
For $T>T_H$, the relevant background (\ref{background}) is $G_{\mu \nu}=\delta_{\mu \nu},\psi=\text{const},\phi\phi^*=\frac{\epsilon}{\kappa^2}$, which leads to
\begin{equation}
F(T_H ^+) = \mathcal{L}_{\phi}|_{\text{background}} = -\frac{c_2 \epsilon^2 T^2}{\kappa^2}.
\end{equation}
Recall that $\epsilon^2 T^2\propto(T-T_H)^2$. It follows that $F$ and its first derivative with respect to $T$ are continuous at $T_H$, but $\frac{d^2 F}{dT^2} $ is not, so the Ehernfest criterion is satisfied. The same calculation shows that the second derivative of the Gibbs free energy density is discontinues when keeping the pressure fixed.

Figure 1 depicts the shape of the thermal scalar potential below (dashed) and above (solid) $T_H$.

\begin{figure}[h]
	\begin{center}
	\includegraphics[scale=0.7]{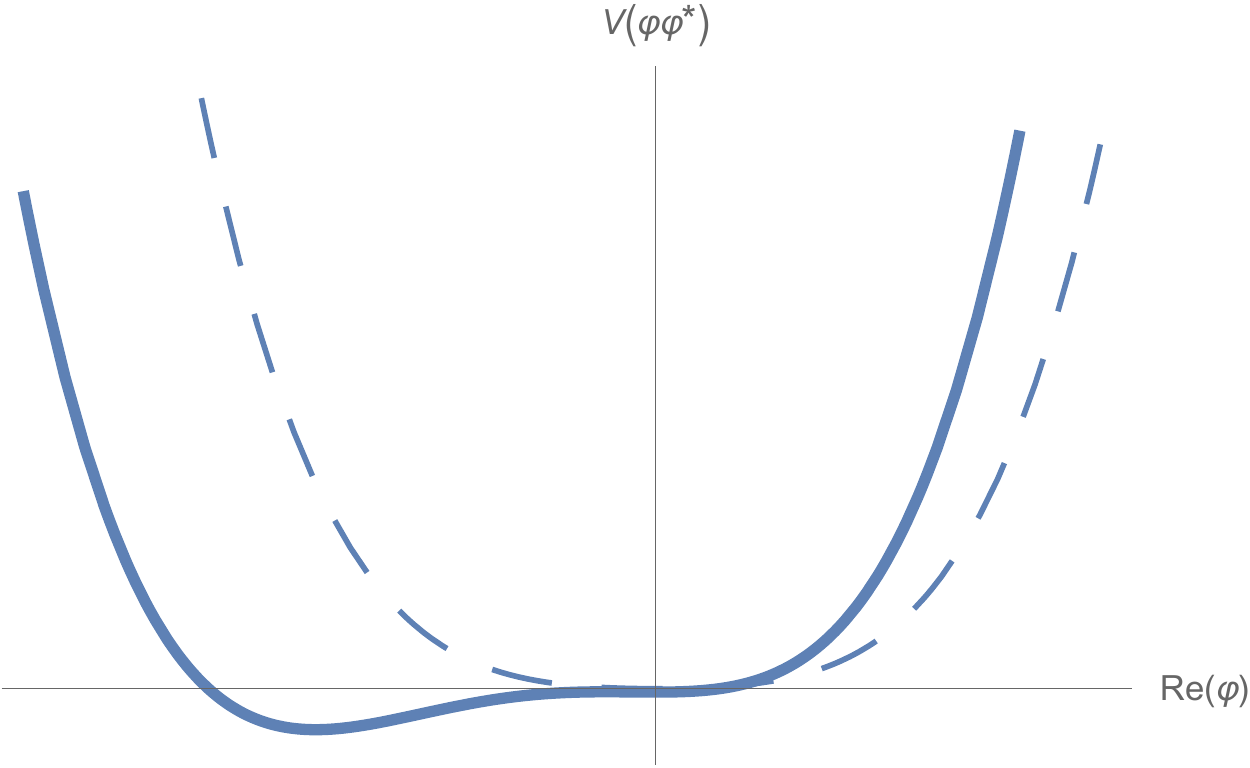}
	\caption{Shown is the thermal scalar effective potential $V(\varphi\varphi^*)$ for $T<T_H$ (dashed) and for $T>T_H$ (solid), as a function of $\text{Re}(\varphi)$ with $\text{Im}(\varphi)$ set to $0$. Below $T_H$, both the mass term and the quartic coupling are positive. Above $T_H$,  the mass term is negative, leading to a mexican-hat tilted potential.}
	\end{center}
\end{figure}

\newpage
\section{Conclusions and Discussion}

We constructed an effective action for closed interacting strings in a thermal state at temperatures close to the Hagedorn temperature by calculating closed string S-matrix elements. It was shown that the thermal scalar condenses above the critical temperature $T_{\text critical}=T_H$. The condensate  background  was shown to correspond to a particular 2D WS CFT. Furthermore, the effective action can be rewritten in variables in which it exhibits conformal symmetry. We also concluded that the Hagedorn transition is a smooth crossover in the bosonic string and a second-order transition for the type II superstring. 

We expect that the heterotic string behaves similarly to the superstring provided that the quartic coupling is still positive. While we made tree-level calculations, considering one-loop effects would not change in a qualitative way the scenario that we described above. First, the thermal scalar self-energy induces an order $O(g_s ^2)$ shift in the Hagedorn temperature (recall that $g_s$ is the dimensionless string coupling constant). Second, masses for the dilaton and antisymmetric tensor are radiatively produced \cite{Panda}. Then, it is possible to integrate them out, in which case the effective action becomes a functional of the thermal scalar and the graviton only.

The EFT that we presented exhibits the phenomenon of symmetry non-restoration at high temperatures \cite{WeinbergSymmetry,Orloff,Pinto}, see also \cite{Chaudhuri} for a recent discussion. The $U(1)$ symmetry that exists below the Hagedorn temperature in the EFT is spontaneously broken above this temperature. Given the expectation that thermal T-duality is spontaneously broken \cite{AW}, the duality would not restore the symmetry.

Our results confirm and extend the qualitative analysis in \cite{dSHigh,CFT4dS} and fix the string-theory-dependent coefficients of the EFT. We expect that it will be possible to use our EFT to strengthen the proposed correspondence between closed strings in the Hagedorn phase and asymptotically de Sitter (dS) space \cite{dSHigh,CFT4dS} and use the EFT to formulate a novel class of string theoretic models of the early Universe.

With appropriate modification to the background, we hope that it will be possible to describe blackholes, extending the ideas about the string BH correspondence \cite{HP,HP1,Damour:1999aw} as proposed in \cite{polymer,emerge} to include also blackholes of sizes that are parametrically larger than the string scale.

\section*{Acknowledgments}

We thank Stefano Giaccari, Sunny Itzhaki, Joey Medved, Kostas Skenderis for useful discussions.

\appendix
\section{Derivation of equation~(16)}

In this appendix we provide additional details on the derivation of Eq.~(\ref{psiphiphi*2}). The equations below are taken or inferred from Polchinski's volume I \cite{PolchinskiI}.
First, the ghost correlation function can be found from Eq.~(6.3.5) of \cite{PolchinskiI}:
\begin{equation}
\langle c\tilde{c}(z_1) c\tilde{c}(z_2)c\tilde{c}(z_3)\rangle = |z_{12}|^2 |z_{13}|^2 |z_{23}|^2 ~,~ z_{ij}= z_i-z_j.
\end{equation}
Second, Eqs.~(8.2.24) and (8.2.25) of \cite{PolchinskiI} imply that the following factors appear in the S-matrix element:
\begin{equation}
2\pi R \delta_{\sum_i n_i ,0} \delta_{\sum_i w_i ,0} \prod_{i,j=1~i<j} ^3 z_{ij} ^{\frac{\alpha'}{2}k_{iL}\cdot k_{jL}}\bar{z}_{ij} ^{\frac{\alpha'}{2}k_{iR}\cdot k_{jR}}.
\end{equation}
Third, Eqs.~(6.2.18-6.2.20) of \cite{PolchinskiI} deal with the contractions of the derivatives of $X$ with exponentials, which give:
\begin{equation}
\left(-i\frac{\alpha'}{2}\right)^2 \left( \frac{k_{1L} ^{\mu}}{z_{31}}+ \frac{k_{2L} ^{\mu}}{z_{32}}\right)\left( \frac{k_{1R} ^{\nu}}{\bar{z}_{31}}+ \frac{k_{2R} ^{\nu}}{\bar{z}_{32}}\right).
\end{equation}
Fourth, the analog of Eq.~(6.6.8) of \cite{PolchinskiI} with one compact dimension implies that the following normalization factor takes place in the calculation:
\begin{equation}
C_{S_2} = \frac{8\pi}{\alpha' (g_c')^2}.
\end{equation}
Finally, the momentum in the non-compact dimensions is conserved due to:
\begin{equation}
(2\pi)^{d} \delta^{d} ({k}_1 ^{\perp} + {k}_2 ^{\perp} + {k}_3 ^{\perp}).
\end{equation}
Gathering the last five factors yields Eq.~(\ref{psiphiphi*2}).

\end{document}